\documentclass[twoside,notoc]{JHEP3}
\pdfoutput=1
\usepackage{epsfig,multicol,slashed}
\usepackage{delarray,amsmath,amssymb,bbm}

\newcommand{\atque}{and\ }
\newcommand{\BY}[1]{#1,}

\newcommand{\nn}{\nonumber}
\newcommand{\beq} {\begin{equation}}
\newcommand{\eeq} {\end{equation}}
\newcommand{\beqa} {\begin{eqnarray}}
\newcommand{\eeqa} {\end{eqnarray}}

\newcommand{\ie}{{\it i.e.}}
\newcommand{\eg}{{\it e.g.}}

\newcommand{\as}{{\alpha_s}}

\newcommand{\eq}[1]{(\ref{#1})}

\newcommand{\inv}[1]{\frac{1}{#1}}
\newcommand{\ket}[1]{\vert{#1}\rangle}
\newcommand{\bra}[1]{\langle{#1}\vert}

\newcommand{\bs}[1]{\boldsymbol{#1}}

\newcommand{\mA}{\mathcal{A}}

\newcommand{\pv}{{\bs{p}}}
\newcommand{\kv}{{\bs{k}_\perp}}
\newcommand{\qv}{{\bs{q}_\perp}}
\newcommand{\bv}{{\bs{b}}}
\newcommand{\ev}{{\bs{e}}}

\newcommand{\qu}{{\rm q}}

\newcommand{\halft}{{\textstyle \frac{1}{2}}}

\title{\center{Measuring transverse size with virtual photons\thanks{Talk at the Third International Workshop on Transverse Polarization Phenomena in Hard Scattering (Transversity 2011), in Veli Lo$\breve{s}$inj, Croatia, 29 August - 2 September 2011.}}}
\author{Paul Hoyer\\
              Department of Physics and Helsinki Institute of
              Physics\\
              POB 64, FIN-00014 University of Helsinki, Finland \\
              }

\abstract{Fourier transforming the virtual photon transverse momentum in $\gamma^*(\qv)+\nobreak N \to f$ processes allows new insight into hadron dynamics as a function of impact parameter $\bv$. I discuss how previous analyses of charge density based on elastic and transition form factors ($f=N,N^*$) can be generalized to any multi-hadron final state ($f=\pi N, \pi\pi N, \bar D \Lambda_c,\ldots$). The $\bv$-distribution determines the transverse positions of the quarks that the photon couples to, and can be studied as a function of multiplicity, the relative transverse momenta, quark masses and polarization. The method requires no factorization nor leading twist approximation. Data with spacelike photon virtualities in the range $0 \leq Q \leq Q_{max}$ provides a resolution $\Delta b \gtrsim 1/Q_{max}$ in impact parameter.}

\begin{document}

\section{Charge density from elastic form factors}

The quark density of a hadron in transverse (impact parameter) space $\bv$ is given by a {\it two}-dimensional Fourier transform of the electromagnetic form factor of the hadron \cite{Soper:1976jc,Burkardt:2000za,Diehl:2002he,Ralston:2001xs}. For a proton with helicity $\lambda$ the density distribution is
\beqa\label{rho0}
\rho_{\qu/N}(\bv) &\equiv& \int \frac{d^2 \qv}{(2 \pi)^2} \,  
e^{-i \, \qv \cdot \bv} \, \frac{1}{2 P^+}   
\bra{P^+, \halft\qv, \lambda}\, j^+(0) \,\ket{ 
P^+, -\halft\qv, \lambda}\nn\\[2mm]
&=& \int_0^\infty \frac{d Q}{2 \pi}\, Q \, J_0(b \, Q) F_1(Q^2)
\eeqa
where $F_1(Q^2)$ is the Dirac form factor. It is not immediately obvious why precisely this definition corresponds to a density. Until recently it was in fact common to define the charge density in terms of a {\it three}-dimensional Fourier transform. Such a definition is, however, not compatible with relativistic effects. Quarks in the proton move with nearly the speed of light, $v_\qu \simeq c$, so a photon cannot give a sharp picture of the charge distribution at an instant of time. However, the {\it transverse} velocity $v_{\qu}^\perp = p_\qu^\perp/E_\qu$ decreases with the energy $E_\qu$ of the quark. In the Infinite Momentum Frame (IMF) $v_{\qu}^\perp = 0$ and a high resolution picture of the charge distribution can be obtained in the {\it transverse plane}, \ie, as a function of impact parameter as in \eq{rho0}.
Formally, the IMF is equivalent to quantization at equal Light-Front (LF) time, $x^+=t+z$. More intuitively, a photon moving along the negative $z$-axis interacts at fixed $x^+$. The Fourier transform in \eq{rho0} is in fact defined in a frame with photon momentum $q^+=0$.
It is also important that the matrix element in \eq{rho0} involves only the $j^+$ component of the quark current, and that the initial and final states have opposite transverse momenta. 

To see why $\rho(\bv)$ merits being viewed as a charge density one needs to expand the hadron $h$ state in terms of its quark and gluon Fock components taken at equal $x^+$ \cite{Brodsky:1989pv},
\begin{eqnarray}\label{fock}
\ket{P^+,\bs{P}_\perp,\lambda}_{x^+=0}^h  &=&  \sum_{n,\lambda_{i}}\,\prod_{i=1}^{n}\Bigl[\int_{0}^{1}\frac{dx_{i}}{\sqrt{x_{i}}}\int\frac{d^{2}\kv_{i}}{16\pi^{3}}\Bigr]16\pi^{3}\delta(1-\sum_{i} x_{i})\,\delta^{(2)}(\sum_{i} \kv_{i})\nn \\[.2cm]
&\times& \psi_{n}^h(x_{i},\kv_{i},\lambda_{i})\,\ket{n;\, x_{i}P^{+},x_{i}\bs{P}_\perp+\kv_{i},\lambda_{i}}
\end{eqnarray}
For a proton the Fock states $n$ would include $\ket{uud},\ket{uudg},\ket{uudu\bar u},\ldots$, the infinite and complete sum of all quark and gluon states, integrated over the longitudinal momentum fraction $x_i$ and the relative transverse momentum $\kv_{i}$ of each parton\footnote{This is a formally exact expansion, but possible contributions from partons with $x_i=0$ (zero-modes) will be neglected.}. The unique property of this LF expansion is that the wave functions $\psi_{n}^h(x_{i},\kv_{i},\lambda_{i})$ do not depend on the hadron momentum $P^+,\bs{P}_\perp$. Hence the {\em same} wave functions $\psi_n^h$ describe the initial and final states in \eq{rho0}. Each parton $i$ carries a share $x_i$ of the parent hadron's longitudinal and transverse momentum, and a relative transverse momentum $\kv_{i}$.

When the initial and final states in \eq{rho0} are expanded in their Fock states \eq{fock}, the impact parameter distribution of a quark $\qu$ is found to be \cite{Diehl:2002he}
\beqa\label{rho0expr}
\rho_{\qu/h}(\bv) &=& \sum_{n,\lambda_{i}}\Bigl[\prod_{i=1}^{n}\int dx_{i}\int 4\pi d^{2}\bv_{i}\Bigr]\delta(1-\sum_{i} x_{i})\frac{1}{4\pi}\delta^{(2)}(\sum_{i} x_{i}\bv_{i})\nn \\[3mm]
 & \times & |\psi_{n}^h(x_{i},\bv_{i},\lambda_{i})|^2\, \sum_k e_{k}\,\delta^{(2)}(\bv-\bv_{k})\, 
\eeqa
where the wave functions $\psi_{n}^h(x_{i},\bv_{i},\lambda_{i})$ are related to the momentum space wave functions $\psi_{n}^h(x_{i},\kv_{i},\lambda_{i})$ of \eq{fock} by standard (two-dimensional) Fourier transforms of the $\kv_{i}$. Thus $\rho_{\qu/h}(\bv)$ indeed is a charge density: the probability that there is a quark $k$ in the hadron at impact parameter $\bv_k=\bv$ (relative to the parent hadron), weighted by its charge $e_k$.

The usual parton distributions $f_{\qu/h}(x,Q^2)$ measured in hard inclusive processes may likewise be expressed in terms of the LF wave functions,
\beqa\label{fq1}
f_{\qu/h}(x,\mu^2)  &=&  \sum_{n,\lambda_{i}} \Bigl[\prod_{i=1}^{n}\int_0^1dx_{i}\int^{k_\perp<\mu}\frac{d^2\kv_{i}}{16\pi^{3}}\Bigr]16\pi^{3}\delta(1-\sum_{i} x_{i})\,\delta^{(2)}(\sum_{i} \kv_{i}) \nn\\[2mm]
 &\times&  |\psi_{n}^h(x_i,\kv_{i},\lambda_{i})|^2\, \sum_k \,\delta(x_k-x)
\eeqa
\EPSFIGURE[h]{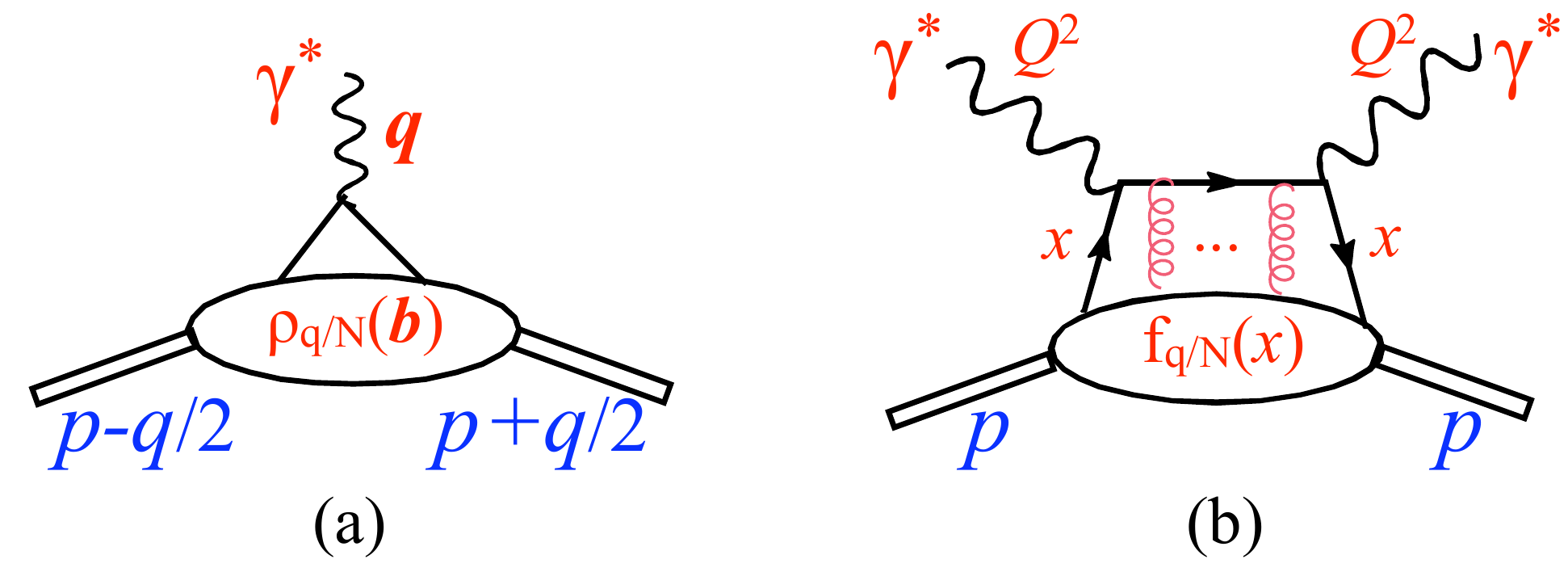,width=.8\columnwidth}
{(a) The elastic proton form factor measured by the $ep \to ep$ scattering amplitude. In the frame $q^+=0$ its Fourier transform \eq{rho0} over $\qv$ gives the charge density in impact parameter. (b) The quark distribution in longitudinal momentum $f_{\qu/h}(x,Q^2)$ \eq{fq1} may be obtained via QCD factorization in the $Q^2 \to \infty$ limit from Deep Inelastic Scattering, $ep \to eX$.}

\noindent It is apparent that the quark distributions in impact parameter \eq{rho0expr} (Fig.~1a) and in longitudinal momentum fraction \eq{fq1} (Fig.~1b) give similar and complementary information on hadron structure. However, there are also important differences. First of all, the expression \eq{fq1} is not exact due to the neglect of the ``Wilson line'', indicated by the vertical Coulomb gluon exchanges in Fig.~1b. This contribution arises from the rescattering of the struck quark in the color field of the hadron and adds coherently to the bound state wave function. There is no Wilson line in the form factor since it has only one photon vertex. Hence the expression \eq{rho0expr} of the impact parameter distribution $\rho_{\qu/h}(\bv)$ in terms of the LF target wave functions is formally exact. 

Parton distributions like $f_{\qu/h}(x,\mu^2)$ are obtained from hard inclusive scattering through QCD factorization at leading power (``twist'') in the $Q^2 \to \infty$ limit, and depend on the factorization scale $\mu$. Corrections to the hard subprocess of higher order in $\as$ must be taken into account in the extraction of $f_{\qu/h}(x,\mu^2)$ from data. Conversely, the expression \eq{rho0expr} for the impact parameter distribution only has corrections from higher orders in the QED coupling $\alpha$. Gluon corrections to the photon vertex are included in the sum over Fock states $n$\footnote{The renormalization of vertex corrections would introduce a scale if the charge density were factorized into a photon vertex and wave functions. However, such scale dependence is absent in the measurable density \eq{rho0expr}. I am grateful for a discussion with Jian-Wei Qiu on this point.}. Data in the whole range of $Q^2$ is used, in fact the integral in \eq{rho0} is over all spacelike $q^2=-\qv^2= -Q^2 \le 0$. A finite range $0 \leq Q \leq Q_{max}=q_{\perp max}$  limits the resolution in impact parameter to $\Delta b \gtrsim 1/Q_{max}$.

\section{Charge distribution of inelastic processes}

The analysis described above has been applied to data on elastic ($\gamma^* N \to N$) and transition ($\gamma^* N \to N^*$) form factors \cite{Miller:2007uy,Carlson:2007xd}. The sum over initial and final Fock states in \eq{rho0expr} remains diagonal for transition form factors, but the product of wave functions ${\psi_{n}^{N^*}}(x_{i},\bv_{i},\lambda_{i})^*\psi_{n}^{N}(x_{i},\bv_{i},\lambda_{i})$ is no longer positive definite. The impact parameter distribution nevertheless reflects the transverse positions of the quarks which couple to the virtual photon.

Here I shall describe the generalization of the impact parameter analysis to any process $\gamma^* i \to f$, where $i$ and $f$ are arbitrary (multi-)hadron states \cite{Hoyer:2011ux}. The possibility to study how the impact parameter distribution depends on the type and relative momenta of the produced hadrons allows new insight into hadron dynamics. I shall use the process $\gamma^* N(p) \to \pi(p_1) N(p_2)$ to illustrate the procedure.

The Fock expansion \eq{fock} is valid for any hadronic state, and thus also for $\ket{\pi(p_1) N(p_2)}$. However, we need to make sure that $\pi N$ states with different total momenta $p_f=p+q=p_1+p_2$ are described by the same LF wave functions. The obvious guess is to parametrize the hadron momenta in the same way as the parton momenta in \eq{fock}:
\beq\label{relmom}
\begin{array}{ll}
p_1^+ = xp_f^+$\hspace{2cm}$ & \pv_{1\perp}=x\pv_{f\perp}+\kv \\[2mm] p_2^+ = (1-x)p_f^+ & \pv_{2\perp}=(1-x)\pv_{f\perp}-\kv
\end{array}
\eeq
We may specify the $\pi N$ state using a wave function $\Psi^f(x,\kv)$ of our choice,
\beq\label{pinpt}
\ket{\pi N(p_f^+,\pv_{f\perp};\Psi^f)}\equiv\int_0^1 \frac{dx}{\sqrt{x(1-x)}} \int\frac{d^2\kv}{16\pi^3}\,\Psi^f(x,\kv) \ket{\pi(p_1)N(p_2)}
\eeq
where $x$ and $\kv$ are independent of the total $\pi N$ momentum $p_f$.
The asymptotic $\ket{\pi N}$ state \eq{pinpt} then has an LF expansion of standard form, with quark and gluon wave functions given by $\Psi^f$ and the Fock state wave functions $\psi_n^\pi$ and $\psi_n^N$ of \eq{fock}. It should be noted that the wave functions which determine the density distribution are the ones at the time of the photon interaction ($x^+=0$), not those of the asymptotic ($x^+ \to \infty$) $\pi N$ state. The $\ket{\pi N}$ state evolves with $x^+$: the hadrons fly apart at large times and converge toward the photon vertex as $x^+ \to 0$. As the pion and the nucleon get close to each other they start to interact and can form resonances. Hence the $\mA(\gamma^* N \to \pi N)$ amplitude has a dynamical phase, unlike the spacelike form factor $\mA(\gamma^* N \to N)$ which is real. A single hadron in the final state has a stationary time development, \ie, its Fock state wave functions are independent of $x^+$.

The impact parameter analysis of the $\gamma^* N \to \pi N$ transition amplitude is similar to that of form factors. In the frame where
\beqa\label{frame}
p=(p^+,p^-,-\halft \qv)\,;\hspace{1cm}
q= (0^+,q^-,\qv)\,;\hspace{1cm}
p_f=(p^+,p^-+q^-,\halft \qv)
\eeqa
the Fourier transform of the $j^+$ current matrix element can be expressed as a diagonal sum over Fock states,
\beqa\label{fta} \hspace{.5cm}
\mA_{fN}(\bv) &\equiv&\int\frac{d^2\qv}{(2\pi)^2}e^{-i\qv\cdot\bv}\inv{2p^+}\bra{f(p_f)}j^+(0)\ket{N(p)}
=\inv{4\pi}\sum_{n}\Bigl[\prod_{i=1}^{n}\int_0^1 dx_{i}\int 4\pi d^{2}\bv_{i}\Bigr] \nn
\\[2mm]
&\times&\delta(1-\sum_{i} x_{i})\delta^{2}(\sum_{i} x_{i}\bv_{i})
 {\psi_{n}^{f}}^*(x_{i},\bv_{i})\psi_{n}^{N}(x_{i},\bv_{i})\,\sum_k e_k\delta^{2}(\bv_k-\bv)
\eeqa
where ${\psi_{n}^{f}}^*$ are the LF wave functions of the state \eq{pinpt} at the photon vertex ($x^+=0$). $\mA_{fN}$ gives the impact parameter distribution of the quarks to which the photon couples, when the center-of-momentum of the initial ($N$) and final ($f$) state is at $\sum_{i} x_{i}\bv_{i}=0$. 

Since the amplitude $\bra{f(p_f)}j^+(0)\ket{N(p)}$ has a dynamical phase due to final-state interactions the Fourier transform in \eq{fta} generally requires a partial wave analysis. Alternatively, the Fourier transform of the {\it square} of the amplitude, 
\beq\label{fsq}
\mathcal{S}_{fN}(\bv) \equiv \int\frac{d^2\qv}{(2\pi)^2}e^{-i\qv\cdot\bv}\,\left|\inv{2p^+}\bra{f(p_f)}J^+(0)\ket{N(p)}\right|^2
=\int d^2\bv_q\, \mA_{fN}(\bv_q)\, \mA_{fN}^*(\bv_q-\bv)
\eeq
gives a convolution of the impact parameter distributions \eq{fta}. Now $\bv$ is the difference between the impact parameter of the quark struck in the amplitude and in its complex conjugate. This also gives a measure of the transverse distribution of the active quarks. The quantity $\mathcal{S}_{fN}(\bv)$ has an imaginary part if the squared amplitude in \eq{fsq} is asymmetric for $\qv \to -\qv$. The imaginary part thus measures the azimuthal correlation between $\qv$ and a transverse direction defined, \eg, by a relative transverse momentum between the particles in the final state (such as $\kv$ in \eq{relmom}).

The square of the amplitude is obtained from the measured cross section -- with the caveat that the analysis concerns only the matrix element of the $j^+$ component of the current. This component dominates at high lepton energies, or may be identified via a Rosenbluth separation.

The above method can be illustrated using the Born level QED amplitude for scattering on a muon, $\gamma^*(q)+ \mu(p) \to \mu(p_1)+\gamma(p_2)$,
\beq\label{mugamq}
\mA^{\mu\gamma,+\frac{1}{2}}_{+\frac{1}{2}+1}(\qv)  =  2e\sqrt{x}\biggl\{\frac{\ev_{-}\cdot\kv}{(1-x)^{2}m^{2}+\kv^{2}}-\frac{\ev_{-}\cdot[\kv-(1-x)\qv]}{(1-x)^{2}m^{2}+[\kv-(1-x)\qv]^{2}}\biggr\}
\eeq
where $\ev_{-}\cdot\kv= e^{-i\phi_{k}}|\kv|/\sqrt{2}$. The muons and the photon are taken to have positive helicities, the final state momenta are parametrized as in \eq{relmom} and the wave function $\Psi^f(x,\kv)$ of \eq{pinpt} is a $\delta$-function in $x$ and $\kv$. The Fourier transform \eq{fta} gives
\beq\label{mugamb}
\mA^{\mu\gamma,+\frac{1}{2}}_{+\frac{1}{2}+1}(\bv)  =  2e\sqrt{x}\biggl[\frac{\ev_{-}\cdot\kv}{(1-x)^{2}m^{2}+\kv^{2}}\delta^{2}(\bv)-\frac{i}{2\sqrt{2}\pi}\frac{m\: e^{-i\phi_{b}}}{1-x}K_{1}(mb)\biggr]\exp\left(-i\frac{\kv\cdot\bv}{1-x}\right)
\eeq
In the first term of \eq{mugamq} the virtual photon interacts with the initial muon, and this term contributes to \eq{mugamb} at the impact parameter $\bv=0$ of the target. In the second term the virtual photon interacts with the muon {\it after} the emission of the real photon, and its $\bv$-dependence agrees with the known $\mu \to \mu+\gamma$ QED wave function. 

If the Fourier transform is taken of the square of the QED amplitude \eq{mugamq} as in \eq{fsq} the result is
\beqa\label{sexp}
\mathcal{S}^{\mu\gamma,+\frac{1}{2}}_{+\frac{1}{2}+1}(\bv;x,\kv) &=&  
4e^{2}x\biggl\{\frac{\kv^{2}/2}{[(1-x)^{2}m^{2}+\kv^{2}]^{2}}\delta^{(2)}(\bv) - im\frac{|\kv|\cos(\phi_{b}-\phi_{k})}{(1-x)^{2}m^{2}+\kv^{2}}\,\frac{K_{1}(mb)}{2\pi(1-x)}\nn\\[2mm]
 &+& \frac{K_{0}(mb)-\halft mb\: K_{1}(mb)}{4\pi(1-x)^{2}}\biggr\}\exp\Big(-i\frac{\kv\cdot\bv}{1-x}\Big)  
\eeqa
The three terms within \{\ \} correspond, respectively, to the virtual photon interacting ({\it i}) with the initial muon in both $\mA^{\mu\gamma}$ and $\big(\mA^{\mu\gamma}\big)^*$, ({\it ii}) once with the intial and once with the final muon, and ({\it iii}) twice with the final muon. The imaginary part can be seen to arise from the angular correlation between the lepton scattering plane (defined by $\bv$) and the relative transverse momentum $\kv$ in the final state.

\section{Discussion}

The analysis presented here generalizes previous work on charge densities of (transition) form factors. Being applicable to any final state in $\gamma^* N \to f$ it opens up a new window on the dynamics of lepton-nucleon scattering. Data at all spacelike photon virtualities $0 \leq Q \leq Q_{max}$ are used, with an expected impact parameter resolution $\Delta b \gtrsim 1/Q_{max}$. The absence of a QCD factorization removes uncertainties related to the leading twist approximation and the factorization scale. The possibilities to study how the impact parameter distribution depends on properties of the final state (multiplicity, relative momenta, quark masses, \ldots) can give new insight into hadron dynamics. For example, the dimensional scaling observed \cite{Bochna:1998ca} in deuteron photodisintegration, $\gamma d \to p n$ at $\theta_{CM} = 90^\circ$, suggests transversally compact configurations of the deuteron and nucleons. A Fourier transform of the electroproduction process, $\gamma^* d \to p n$, can reveal the transverse distribution of the active quarks.

The absence of QCD factorization also implies less predictions. The $Q^2$-dependence of the quark distributions $f_{\qu/N}(x,Q^2)$ measured in DIS can be calculated, and the universality of the distributions tested in other hard processes. The LF wave functions $\psi_n$ which determine the impact parameter distribution in \eq{fta} are also universal, but are more difficult to reconstruct from measured data. The impact parameter analysis discussed here thus is complementary to the traditional analyses of hard inclusive (and exclusive) scattering processes.

\acknowledgments
I wish to thank the organizers of the Transversity 2011 workshop for their kind invitation. The Magnus Ehrnrooth foundation provided travel support.

\end{document}